# System-size scaling of Boltzmann and alternate Gibbs entropies

Jose M. G. Vilar[1,2] and J. Miguel Rubi[3,4]

[1]Biophysics Unit (CSIC-UPV/EHU), University of the Basque Country, Box 644, 48080 Bilbao, Spain. [2]IKERBASQUE, Basque Foundation for Science, 48011 Bilbao, Spain. [3]Departament de Fisica Fonamental, Universitat de Barcelona, Diagonal 647, 08028 Barcelona, Spain. [4]Department of Chemistry, Imperial College London, SW7 2AZ, London, UK.

## Abstract

It has recurrently been proposed that the Boltzmann textbook definition of entropy $S(E) = k \ln \Omega(E)$ in terms of the number of microstates $\Omega(E)$ with energy $E$ should be replaced by the expression $S_G(E) = k \ln \sum_{E'<E} \Omega(E')$ examined by Gibbs. Here, we show that $S_G$ either is equivalent to $S$ in the macroscopic limit or becomes independent of the energy exponentially fast as the system size increases. The resulting exponential scaling makes the realistic use of $S_G$ unfeasible and leads in general to temperatures that are inconsistent with the notions of hot and cold.

## Introduction

The well-established textbook definition of entropy $S(E) = k \ln \Omega(E)$ in terms of the number of microstates $\Omega(E)$ with energy $E$ was introduced by Boltzmann, reformulated by Plank in its present form, and subsequently generalized by Gibbs through the ensemble approach [1]. Since then, this formula has been the cornerstone of statistical physics. In his seminal monograph [2], Gibbs also explored the properties of the continuous phase space counterpart of the alternative definition $S_G(E) = k \ln \sum_{E'<E} \Omega(E')$, mainly as a calculation device because of its equivalence to $S$ in the macroscopic limit of the systems of interest at the time. This macroscopic equivalence has been exploited backwards recurrently over the years to propose that the standard definition of entropy $S$ should be replaced by $S_G$ [3-6]. Specially prominent has been the use of $S_G$ to negate the existence of negative temperatures [5,6], which might seem unappealing and counterintuitive but which are inevitable in systems with bounded energy spectra [7]. In these types of systems, which range from nuclear spins [8,9] to trapped ultracold atoms [10,11], the number of microstates $\Omega(E)$ has a maximum for finite energies and $S$ and $S_G$ do not coincide with each other in the thermodynamic limit.

Here, we show that $S_G$, in contrast to $S$, ceases to be a function of the energy for decreasing $\Omega(E)$ in the macroscopic limit and that it does so exponentially fast. Such exponential dependence makes meaningful use of $S_G$ unfeasible not only for macroscopic systems but also for small systems with over tens of elements and leads to temperatures that are inconsistent with the notions of hot and cold.

**Results**

The fact that $S_G$ ceases to be a function of the energy for decreasing $\Omega(E)$ in the macroscopic limit follows straightforwardly from the maximum-term approach [12], which shows that the logarithm of a sum can be approximated by the logarithm of the maximum term. It leads to $S_G(E) = k \ln \Omega(E^*)$ in the macroscopic limit, where $E^*(\leq E)$ is the energy that maximizes the number of microstates. This result can be worked out explicitly by considering the energy levels indexed by $u$ from $u = 0$ to $u = U$ so that $E_u < E_{u+1}$ and $E_U = E$. The value of the sum $\sigma(E) = \sum_{u=0}^{U} \Omega(E_u)$ is greater than the value of the largest term, $\Omega(E^*)$, and smaller than the number of terms, $1+U$, times the value of the largest term. In mathematical terms, these conditions are expressed as $\Omega(E^*) \leq \sigma(E) \leq (1+U)\Omega(E^*)$. Taking logarithms and multiplying by $k/N$ gives $S(E^*)/N \leq S_G(E)/N \leq S(E^*)/N + [k \ln(1+U)]/N$. Since the number of energy levels grows subexponentially with the system size, these bounds imply that $S_G(E) = S(E^*)$ for large $N$. Therefore, both definitions of entropy are the same in the macroscopic limit if the number of microstates increases continuously with the energy since $E^* = E$ for all $E$. However, when the number of microstates decreases with the energy, $S_G$ becomes constant for all $E > E^*$.

This general result is illustrated explicitly by the prototypical ensemble of $N$ two-level units with energies 0 and $\varepsilon$ and total energy $E = \varepsilon U$, where $U$ is the number of units in the higher energy level [13]. In this case, the number of microstates is given by $\Omega(E) = \dfrac{N!}{(N - E/\varepsilon)!(E/\varepsilon)!}$, which leads to $S_G = k \ln \sum_{u=0}^{E/\varepsilon} \dfrac{N!}{(N-u)!u!}$. As the system size increases, $S_G$ looses its dependence on $E$ for $E > \varepsilon N/2$ (Fig. 1).

A fundamental question for the validity of $S_G$ as feasible thermodynamic quantity is how fast $S_G$ ceases to be a function of the energy for $E > E^*$. Explicitly, the key question is whether $S_G$ can physically provide information on the thermal properties of the system for large but finite $N$.

The loss of thermal information can be quantified explicitly through the difference $\Delta S_G = S_G(E_U) - S_G(E_{U-1})$, which indicates how $S_G$ changes between two contiguous energy levels. This quantity is related to the associated temperature through $T_G = (E_U - E_{U-1})/\Delta S_G$, the discrete counterpart of the macroscopic expression $T_G = (\partial S_G / \partial E)^{-1}$. Using $\Omega(E) = e^{S(E)/k}$ in $S_G$ leads to

$$\Delta S_G = k \ln(1 + e^{S(E_U)/k} / \textstyle\sum_{u=0}^{U-1} e^{S(E_u)/k}) .$$

An upper bound that indicates explicitly how fast $\Delta S_G$ goes to zero as $N$ increases for $E > E^*$ can be obtained by making use of two inequalities. The first one, $\ln(1+x) < x$, leads to

$$\Delta S_G < k e^{S(E_U)/k} / \textstyle\sum_{u=0}^{U-1} e^{S(E_u)/k} ,$$

which together with the second one, $\sum_{u=0}^{U-1} e^{S(E_u)/k} > e^{S(E^*)/k}$, valid for $E = E_U > E^*$ or equivalently $E_{U-1} \geq E^*$, results in

$$\Delta S_G < k e^{S(E)/k - S(E^*)/k}.$$

Since the entropy $S$ is an extensive quantity and $S(E) < S(E^*)$, this result explicitly shows that $S_G$ ceases to be a function of the energy exponentially fast for $E > E^*$ as the system size increases. As a result, $S_G$ cannot physically provide feasible information on the thermal properties of the system for $E > E^*$.

The resulting temperature $T_G = (E_U - E_{U-1}) / \Delta S_G$, in turn, is not consistent with an intensive quantity, as required by thermodynamics, but instead it growths exponentially with the system size indefinitely for all $E = E_U > E^*$. This exponential behavior makes impossible a meaningful association of $T_G$ to a physical quantity since doubling the system size, for instance, increases $T_G$ several orders of magnitude even for relatively small systems far below the macroscopic or mesoscopic limit.

The prototypical ensemble of $N$ two-level units discussed previously clearly illustrates the implications of this pathological behavior. In this case, making use of $\Omega(E) = \frac{N!}{(N - E/\varepsilon)!(E/\varepsilon)!}$ and $\Omega(E^*) = \frac{N!}{[(N/2)!]^2}$ in the formula of the entropy $S$, we obtain

$$\Delta S_G < k \frac{[(N/2)!]^2}{(N-U)!U!}.$$

It is possible to use Stirling's approximation in the previous expression to obtain an approximate bound but using $n! \leq e\, n^{n+1/2} e^{-n}$ in the numerator and $\sqrt{2\pi}\, n^{n+1/2} e^{-n} \leq n!$ in the denominator allows us to obtain the precise bound

$$\Delta S_G < k \frac{(N/2)^{N+1} e^{-N+2}}{2\pi (N-U)^{N-U+1/2} U^{U+1/2} e^{-N}} = k \frac{e^2}{2\pi \sqrt{(1-x)x}} [2(1-x)^{1-x} x^x]^{-N},$$

where $x = U/N$. Consequently, the resulting temperature $T_G = \varepsilon / \Delta S_G$ grows exponentially with the system size as $T_G > \upsilon e^{\alpha N}$ with $\alpha = \ln 2 + (1-x)\ln(1-x) + x\ln x$ and $\upsilon = ke^2 / 2\pi\sqrt{(1-x)x}$ as illustrated in Fig. 2. In this case, doubling the system size from $N = 100$ to $N = 200$ already increases $T_G$ over 5 orders of magnitude for $x = 0.75$.

What happens then when systems of different sizes exchange energy with each other? Consider for instance a system A with $N = 100$ and $U = 55$ and a system B with $N = 10{,}000$ and $U = 5{,}000$. When the two systems are allowed to exchange energy with each other, the energy will be redistributed so that the average energy of each element is the same and heat will flow from system A to system B (Fig. 3). The entropy $S_G$, however, assigns a temperature $T_G = 17.33\varepsilon / k$ to system A, which is lower than the temperature $T_G = 62.67\varepsilon / k$ that it assigns to system B. Therefore, $T_G$ is not consistent with the notions of hot and cold and it would imply heat spontaneously flowing from low to high temperatures.

## Discussion

The definition of entropy is the main cornerstone of statistical physics. Our results have shown that the recurrent proposal $S_G(E) = k \ln \sum_{E'<E} \Omega(E')$ as fundamental entropy cannot faithfully describe physical systems at any scale. Macroscopically, $S_G$ is either identical to $S$ or unrealistically independent of the energy. For finite systems, from small to large, the resulting temperature $T_G$ is not consistent with the notions of hot and cold, implying that heat can spontaneously flow from low to high $T_G$. Thus, our results strongly support the concept of absolute negative temperature as measured experimentally in nuclear spins [8,9] and trapped ultracold atoms [10,11], which has so prominently been contested recently [6].

## Acknowledgments

We thank Roberto Piazza for discussions and for bringing to our attention that the main results of Ref. [6] were obtained previously in Refs. [4,5]. This work was supported by the MINECO under grants FIS2012-38105 (J.M.G.V.) and FIS2011-22603 (J.M.R.).

## Figure legends

**Figure 1. System-size scaling of $S_G$.** The normalized entropy $S_G$ of the ensemble of two-level units is shown as a function of the normalized energy for system sizes $N = 3$, 10, 30, and 100 (from lighter to darker colors).

**Figure 2. Exponential growth of $T_G$ with the system size $N$.** The dimensionless temperature $kT_G / \varepsilon$ of the ensemble of two-level units is shown as a function of the system size $N$ for $x = 0.75$ (continuous black line). The lower bound $k\upsilon e^{\alpha N} / \varepsilon$ is shown as a dashed gray line.

**Figure 3. Inconsistency of $T_G$ with the notions of hot and cold.** The averages over 300 realizations of the time evolution of $U$ for the ensemble of two-level units are shown for systems A ($N = 100$) and B ($N = 10^4$) upon coupling. In each time step, the state of the system is updated by swapping the states of two randomly picked elements in systems A and B. In this case, heat spontaneously flows from system A to system B even though $T_G$ for system A is lower than $T_G$ for system B.

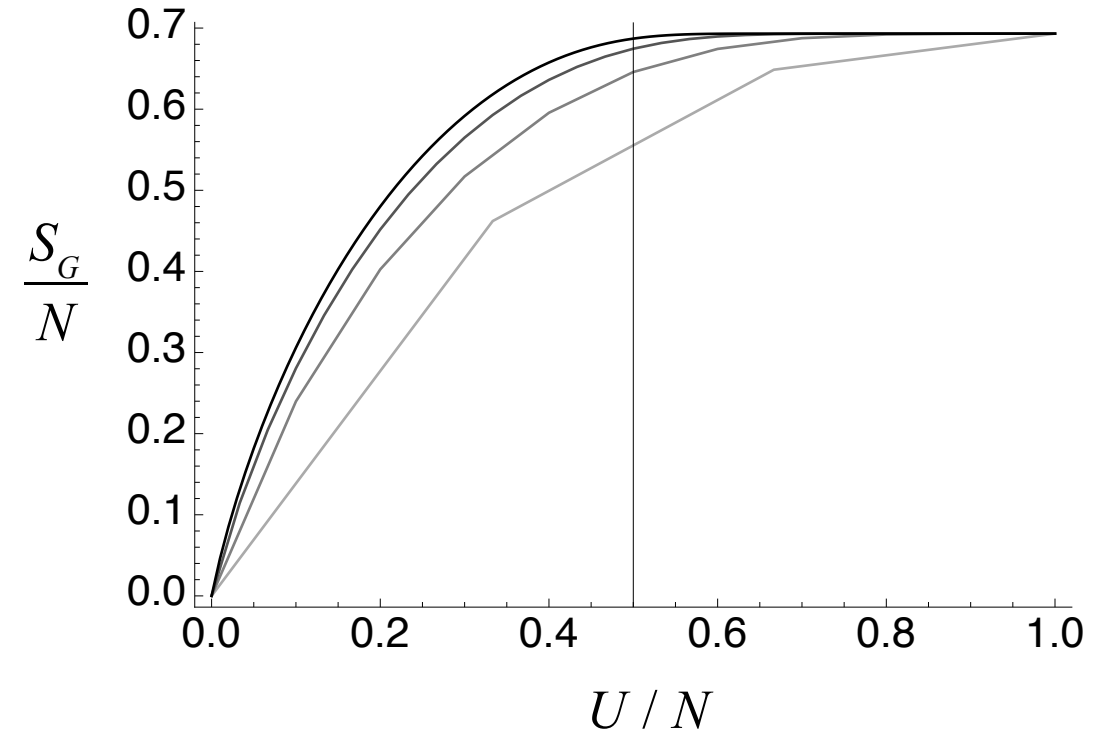

Figure 1

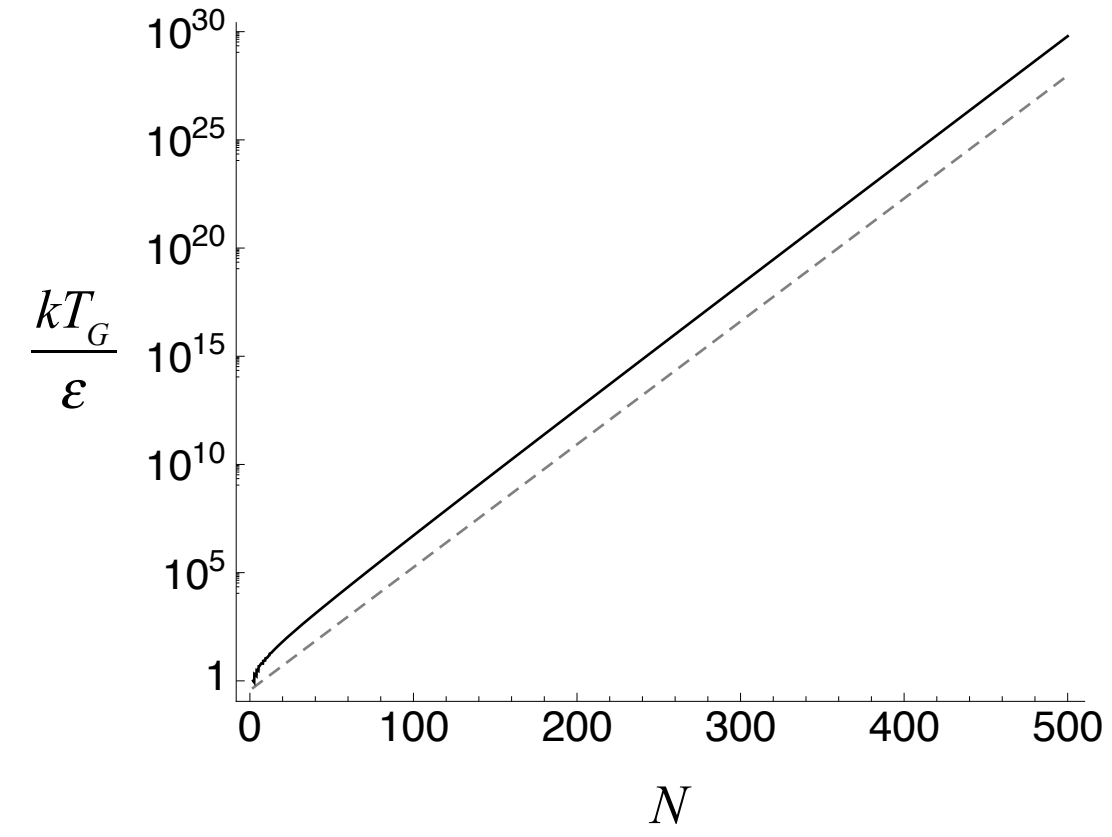

Figure 2

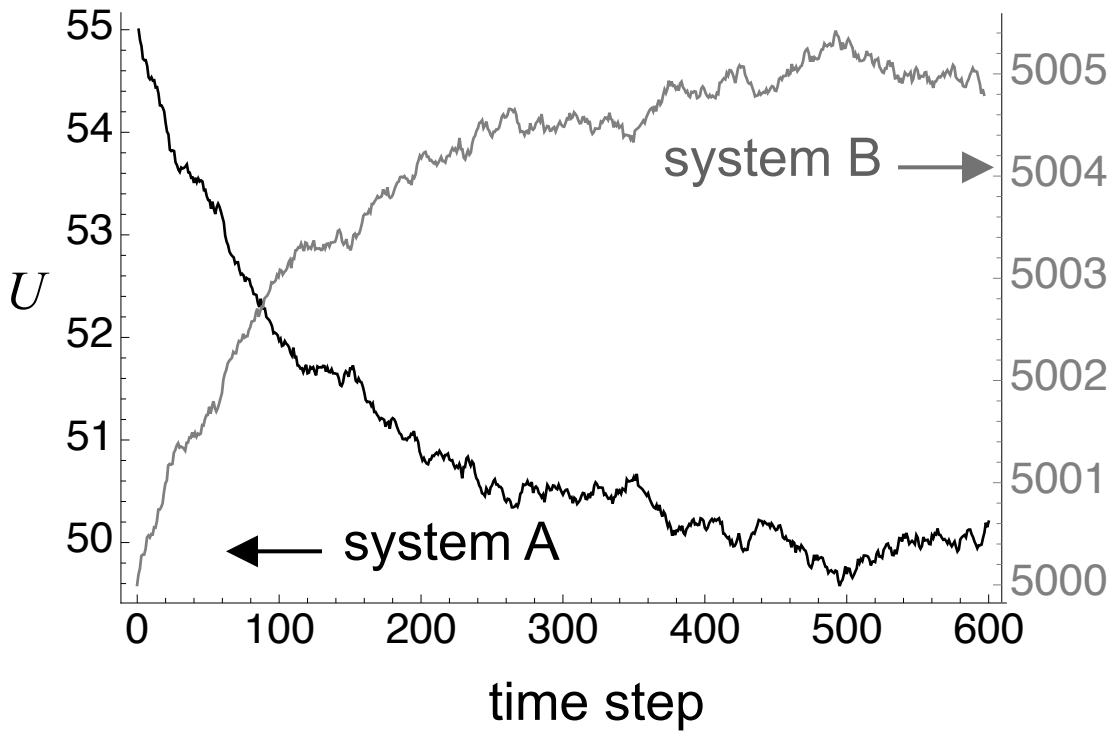

Figure 3